\documentstyle[prl,twocolumn,aps,epsf]{revtex}

\input  epsf
\begin{document}
\twocolumn[\hsize\textwidth\columnwidth\hsize\csname@twocolumnfalse%
\endcsname
\title{Numerical study of relaxation in electron glasses}
\author{
A.\ P\'erez-Garrido, M.\ Ortu\~no,  A. D\'\i az-S\'anchez and E. Cuevas \\
{\ {Departamento de F\'{\i}sica, Universidad de Murcia, Murcia 30.071, Spain}
}}

\maketitle

\begin{abstract}
{
We perform a numerical simulation of energy relaxation in three--dimensional 
electron glasses in the strongly localized regime at finite temperatures. 
We consider systems with no interactions, with long-range Coulomb 
interactions and with short-range interactions,
obtaining a power law relaxation with an exponent of 0.15, which is independent of the
parameters of the problem and of the type of interaction. 
At very long times, we always find an exponential regime whose
characteristic time strongly depends on temperature, system size,
interaction type and localization radius. We extrapolate the longest
relaxation time to macroscopic sizes and, for interacting samples, obtain
values much larger than the measuring time.
We finally study the number of electrons participating in the relaxation
processes of very low energy configurations.}
\end{abstract}
\pacs{PACS numbers: 71.55.Jv, 72.80.Ng, 71.27.+a }
]

\section{Introduction}

Strongly localized systems are characterized by very slow relaxation
rates due to the exponential dependence of the transition rates on
hopping length \cite{CK92,OP97,VO98}. For a wide range of parameters, 
the typical times involved are much larger than the experimental times 
and a glassy behavior is observed. 
Ben--Chorin
{\it et al.} \cite{CK92} reported on non-ergodic transport in Anderson
localized films of indium--oxide and ascribed the phenomena to the
hopping transport in non-equilibrium states. 
Ovadyahu and Pollak \cite{OP97}
performed further experiments on this system that clearly demostrate the
glassy nature of Anderson insulators.
Glassy behavior may be obtained independently of the 
strength of interactions and regardless of their long or short range.
In systems with localized states, long hopping lengths result 
in very long relaxation times. 
However, it is thought  that there are specific features of the glassy
relaxation behavior that indeed depend on the type and strength 
of the interactions involved. If so, relaxation experiments could be an 
adequate tool for studying the strength of interactions. 
There has been no systematic study
of the effects of interactions on the relaxation properties of strongly
localized systems, and in this paper we try to fill this gap as much as 
possible.

Most properties of systems with localized electronic states strongly
depend on interactions. This is especially true for Coulomb glasses where 
interactions are of a long range character. 
The non--equilibrium properties of these systems are affected by dynamic 
correlations in the motion of electrons \cite{PO85}. 
One--particle densities of states or excitations are not enough to
encompass  the whole problem.
To deal with such problems, methods were developed \cite{MP91,ST93,DM98} to 
obtain the low lying states and energies of electron glasses.
The states of the system, their energies and the transition rates between 
them constitute the information needed to compute non-equilibrium properties.
We use this information to study energy relaxation for systems with no 
interactions, with long-range Coulomb interactions and with short-range 
interactions.

In the next section, we describe the model and the numerical procedure
used. In section III, we study the temporal dependence of energy relaxation
and, in section IV, we calculate the largest relaxation time $\tau_2$ and 
its dependence on size and temperature. Finally, in section V, we
present results about the number of electrons participating in low-
energy relaxation processes.

\section{Model and numerical procedure}

We consider three--dimensional systems in the strongly localized regime,
in which quantum overlap
energies, $h$, arising from tunnelling are much smaller than the other
important energies in the problem and are taken into account only to the
lowest contributing order, {\it i.e.},
to zero order for energies and to first order for
transition rates. Spin is neglected since exchange energies are proportional
to $t^2$.
We use the standard tight--binding Coulomb gap Hamiltonian \cite{SE84}:
\begin{equation}
H=\sum_i\epsilon_i n_i +\sum_{i<j} n_in_jV_{ij}\;, \label{hamil}
\end{equation}
where $\epsilon_i$ is the random site energy chosen from a box
distribution with interval $[-W/2, W/2]$. 
For non-interacting systems $V_{ij}=0$, while $V_{ij}=1/r$ for systems with Coulomb 
interactions and $V_{ij}=(0.7/r)^4$ is the potential chosen for short range 
interactions.
The large value of the Hubbard energy is accounted for by disallowing
double occupation of sites. 

We study systems with sizes from 248 to 900 sites 
placed at random (for short range interactions we only
consider systems sizes up to 465 sites), but with a minimum separation
between them, which we choose to be $0.5l_0$ where $l_0=(4\pi N/3)^{-1/3}$
and $N$ is the concentration of sites. 
We take $e^2/l_0$ as our unit of energy and $l_0$ as our unit of distance.
We choose the number of electrons to be equal to half the number of sites.
We use cyclic boundary conditions. 

We use two different numerical algorithms to obtain the ground state and
the lowest energy many--particle configurations of the systems
up to a certain energy.
For short range interactions, 
we employ an algorithm that relaxes the system through certain simultaneous 
$n$--electron transitions \cite{PO97}.
The procedure is repeated for different initial random
configurations of the charges until the configuration of lowest energy is
found ten times. The configurations thus generated were memorized in terms
of site occupation numbers and of energy, whenever this was less
than the highest energy configuration in memory storage.
We complete the set of low--energy configurations by generating all
the states that differ by one-- or two--electron transitions from any
configuration stored. 

For long range interactions, we 
use an algorithm that consists of finding the low-energy many-particle
configurations by means of a three-step algorithm \cite{ADMO98}.
This comprises local search 
\cite{MOPO96,PO97}, thermal cycling \cite{MO97} and construction
of ``neighbouring'' states by local re-arrangements of the charges 
\cite{MOPO96,PO97}. The efficiency of this algorithm is discussed in 
Ref.\ \cite{ADMO98}. In the first step, an initial set, 
${\cal S}$, of metastable low-energy many-particle states is created. We start from 
states chosen at random. These states are relaxed by a local search 
algorithm which ensures stability with respect to excitations from one 
to four sites. In the second step, this set ${\cal S}$ is improved by 
means of the thermal cycling method, which combines the Metropolis and 
local search algorithms. Lastly, the third step completes the set ${\cal S}$ by 
systematical investigations of the surroundings of the states previously 
found.

The transition rate $\omega_{IJ}$ 
between configurations $I$ and $J$ is taken to be
\begin{equation}
\omega_{IJ} =\frac{1}{\tau_0}\exp \left(-2\sum r_{ij}/a\right)
\exp\left(-\frac{E_J-E_I}{kT}\right)
\label{omega}
\end{equation}
for $E_J>E_I$, and without the second exponential for $E_J<E_I$.
In this equation, $\tau_0$ is the inverse phonon frequency, of the order of
$10^{-13}$ s, $a$ is the localization radius, which we take 
equal to $0.3l_0$, and $\sum r_{ij}$ 
is the minimized sum of the hopping lengths of the electrons
participating in the transition.

The relaxation process is governed by the master equation, which in first
order can be written in matrix form as 
${\bf p}(t+\delta t)= {\bf \mathcal{M}\bf p}(t)$, where 
${\bf p}$ is the vector of
occupation probabilities in the configuration space,
and ${\bf \mathcal{M}}$ the matrix of transition
probabilities between states during a time, $\delta t$, given
by \cite{RC94,PO98}:
\begin{equation}
({\bf \mathcal{M}})_{JI}=\left\{
             \begin{array}{ll}
              \omega_{IJ}\delta t&{\rm for}\; I\neq J,\\
               1-\sum_{K\neq I}\omega_{IK}\delta t &{\rm for}\; I=J.
              \end{array} \right.
\end{equation}

We assume that the system initially occupies a set, $\mathcal{K}$, of $m$
configurations with equal probabilities, that is,
$p^{(0)}_K=1/m$ for $K\in \mathcal{K}$, and $p^{(0)}_L=0$ for all other
$L$. The time evolution of  ${\bf p}$ is governed by the eigenvalues
$\lambda_i$ and right eigenvectors $\vec{\phi}_i$ of ${\bf \mathcal{M}}$. 
We will assume that the $\lambda_i$ are arranged in decreasing order.
Rewriting ${\bf p}^{(0)}$ as a linear combination of the $\vec{\phi}_i$, 
the probability vector after $n$ time steps ${\bf p}^{(n)}$ is given by
\begin{equation}{\bf p}^{(n)}=a_1 \vec{\phi}_1 + a_2\vec{\phi}_2\lambda_2^n +
a_3\vec{\phi}_3\lambda_3^n +...\label{infinito}
\end{equation}
where $a_i$ is the $i$-th component of ${\bf p}^{(0)}$ in the basis
$\left\{\vec{\phi}_i\right\}$. At long
times (large $n$), Eq. (\ref{infinito}) approaches equilibrium with time
dependences given by $\lambda_i^n$. Thus, the 
relaxation times are given by 
\begin{equation}
\tau_i={1\over |{\rm ln}\lambda_i|}\label{tiempos}
\label{tau}
\end{equation}
in units of $\delta t$. 
The final state is  $p^{(\infty)}_M=\exp (-E_M/kT)/Z$ for
all $M$, where $E_M$ is the energy of state $M$, and $Z$ is the 
partition function.
Clearly ${\bf p}^{(\infty)}$ is a right eigenvector of ${\bf \mathcal{M}}$ 
with eigenvalue 1, since  ${\bf \mathcal{M}\bf p}^{(\infty)}={\bf p}^{(\infty)}$.
All the other eigenvalues of ${\bf \mathcal{M}}$ are smaller than 1, since
otherwise the system would not tend to the stationary probability
distribution. 
The second largest eigenvalue corresponds to
the largest relaxation time of the system.
The addition of the other eigenvectors to $\phi_1={\bf p}^{(\infty)}$,
transfers ${\bf p}$ from high energy states to low energy states at various rates.

We have developed a renormalization method to be able to properly handle the
huge range of transition rates involved. 
Large values of $\tau_i$ correspond to $\lambda_i$ with values which are very 
close to unity, 
Eq. (\ref{tau}), and a direct calculation of $\tau_i$, in units of
$\delta t$, is strongly limited by the numerical precision of the computer. 
In order to minimize errors, we must choose a $\delta t$ which is
as large as possible, although this  soon yields negative diagonal elements 
of ${\bf \mathcal{M}}$.
We overcome this problem using a renormalization procedure that allows us 
to increase $\delta t$ and to  simultaneously keep  all terms 
of ${\bf \mathcal{M}}$ positive. This procedure forms groups of configurations.
Each group is made up of configurations connected between themselves by 
transition rates which are larger than a critical one. The groups are clusters in local
equilibrium for times greater than the inverse of the critical
transition rate. Firstly, we take a critical transition rate
$\omega_{\rm c}$. Then for each $\omega_{IJ}$ larger than $\omega_{\rm c}$ 
we define a new equilibrium state, $M$, and substitute the original
configurations, $I$ and $J$, by this new state, $M$. 
The transition rates between $M$ and any other configuration $K$ 
($K\neq I,J$) are defined as:
\begin{eqnarray}
&&\omega_{KM}=\omega_{KI}+\omega_{KJ}\\
&&\omega_{MK}=\frac{\omega_{IK}}{1+R_M}+\frac{\omega_{JK}}{1+R_M^{-1}}
\label{rates}
\end{eqnarray}
where $R_M$ is given by:
\begin{equation}
R_M=\frac{\omega_{IJ}}{\omega_{JI}}=\exp \left\{(E_I-E_J)/k_{\rm B}T
\right\}.\label{radio}
\end{equation}
The diagonal matrix elements $\omega_{MM}$ are again equal to 1 minus
the sum of the non-diagonal elements of the column $M$ multiplied by
$\Delta t$.

After the matrix ${\bf \mathcal{M}}$ has been renormalized by the previous
procedure, we can increase 
the time scale to a larger interval $\delta^\prime t= 1/\omega_{\rm c}$. 
With this $\delta^\prime t$ we calculate the new
elements of ${\bf \mathcal{M}}$. The eigenvalues of the transition matrix
will be given now in units of $\delta^\prime t (>\delta t)$.
We have checked the validity of our renormalization procedure with
several samples of small systems where errors are not critical. 
The method minimizes computer errors in the solution of the eigenproblem as
the matrix becomes less ill-conditioned, and allows us to consider large
systems, with matrix elements that differ by many orders of magnitude.

\section{Temporal dependence}

\begin{figure}
\epsfxsize=\hsize
\begin{center}
\leavevmode
\epsfbox{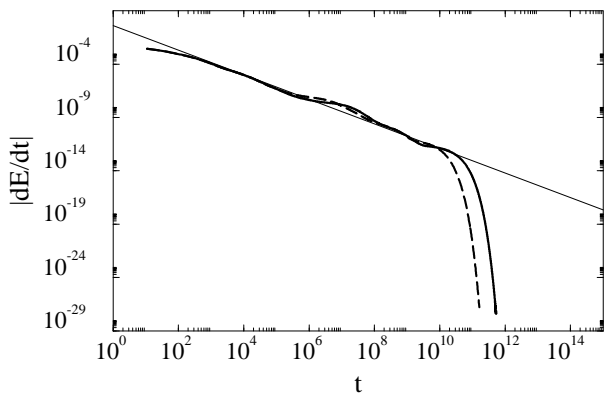}
\end{center}
\caption{Double log$_{10}$ plot of the temporal derivative of the 
relaxation energy versus time for a system with Coulomb interaction,
for $T=0.004$ (solid curve) and 0.005 (dashed curve).
The straight line corresponds to power--law relaxation, and has a 
slope equal to $-1.15$. $t$ is given in units of $\tau_0$.}
\label{t_coulomb}
\end{figure}

We calculate the temporal dependence of the
energy of the system when it relaxes from an initial set of high energy
configurations. At very long times, the longest relaxation process
involved predominates and we see an exponential relaxation. For shorter
times, there is an almost continuous sequence of relaxation times,
which gives rise to a power law relaxation $(E-E_{\rm eq})\propto 
t^{-\alpha}$. 
To obtain the exponent of
this law it is convenient to represent the absolute derivative of the
energy with respect to time. In Fig.\ \ref{t_coulomb} we show
$|{\rm d}E/{\rm d}t|$ versus time (in units of $\tau_0$) 
in a double log$_{10}$ plot for a sample with 
Coulomb interactions and 248 sites. The continuous curve corresponds to
a temperature $T=0.004$, and the dashed curve to $T=0.005$.
The straight line is a fit to the data in the 
non--exponential part of both curves, and its slope is equal to $-1.15$. 
So the power--law exponent for relaxation is $\alpha=0.15$. 
This exponent is basically independent of temperature for all the systems
considered.

\begin{figure}
\epsfxsize=\hsize
\begin{center}
\leavevmode
\epsfbox{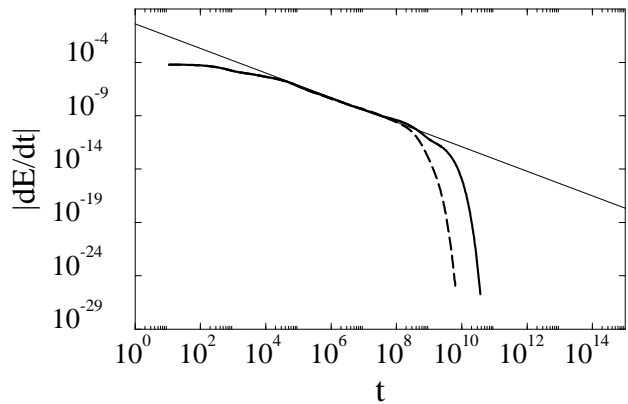}
\end{center}
\caption{Double log$_{10}$ plot of $|{\rm d}E/{\rm d}t|$ versus time 
for a non--interacting system, for $T=0.004$ (solid curve) and 0.005 
(dashed curve). The straight line has a slope equal to $-1.15$.
$t$ is given in units of $\tau_0$.}
\label{t_nothing}
\end{figure}

We have also studied energy relaxation for systems with short--range
interactions and for non--interacting systems. The results for 
short--range interactions are very similar to those for Coulomb 
interactions. The power--law exponent is roughly 0.15 and the largest
relaxation time is of the same order of magnitude as for Coulomb
systems.
In Fig.\ \ref{t_nothing} we show 
$|{\rm d}E/{\rm d}t|$ as a function of time in a double log$_{10}$ plot 
for a  non--interacting system with $N=248$ sites. 
The continuous curve is for $T=0.004$, and the dashed curve for $T=0.005$.
The slope of the straight line is again equal to $-1.15$. 
There are two differences between the results for interacting and for
non--interacting systems. The longest relaxation times are shorter for
the latter, and the power--law regime is not very well defined in the
absence of interactions. Both figures give the rate of relaxation $|{\rm d}E/{\rm d}t|$
at any time. At very small $t$, the interacting systems relax faster
than the non--interacting systems. A possible explanation
of this is that in the excited state of the interacting systems some electrons get
very close to
each other. In the initial stages of relaxation these electrons hop away from electrons
in the nearest neighbors sites, the whole process being very fast.

Several samples have been checked  and in all of them we obtain similar
results to Figs.\ \ref{t_coulomb} and \ref{t_nothing}.  
Two features characterize our relaxation process, the exponent $\alpha$
of the power--law regime and the longest relaxation time.
The exponents $\alpha$ do not appreciably vary from sample to sample,
nor with temperature or with the type of interaction.
On the other hand, the longest relaxation time drastically changes from 
sample to sample and with changes in  temperature, size and the range of interaction.
On average, this time increases 
with the size of the system and with the strength of the interactions. 
In the next section we study the longest relaxation time in detail. 
Now we shall analyze  exponent $\alpha$.

\begin{figure}
\epsfxsize=\hsize
\begin{center}
\leavevmode
\epsfbox{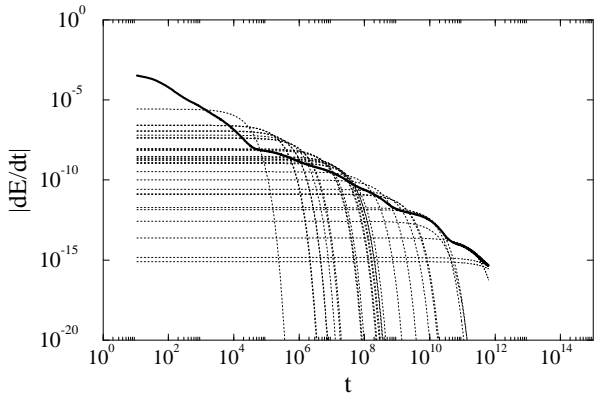}
\end{center}
\caption{Exponential relaxation processes coming from several eigenvalues of 
$\mathcal{M}$ in a double log$_{10}$ plot. A power law arises from the combination of all of them.}
\label{t_several}
\end{figure}

Temporal relaxation can be described as a sum of parallel exponential
relaxation processes, each  with its own different relaxation time,
$\tau_i$. The energy, $E$, of the system can be written as a
function of time, $t$, as follows
\begin{equation}
E(t)=\sum_{i>1}c_i\exp \left( -\frac{t}{\tau_i}\right)+E_{\rm Eq}
\end{equation}
where $c_i$ is the product of the $i$-th component of the initial
occupation vector, $a_i$, and the energy associated to the eigenvector 
$\phi_i$. This energy is the sum of the components of $\phi_i$
multiplied by the corresponding energies.
$E_{\rm Eq}$ is the equilibrium energy, i. e. $E_{\rm Eq}=E(t \longrightarrow\infty )$. 
In Fig.\ \ref{t_several} we plot
$(c_i/\tau_i)\exp(-t/\tau_i)$ for the 30 largest eigenvalues of
$\mathcal{M}$, excluding $\lambda_1=1$, as a function of time for a sample
with Coulomb interactions and of size $N=465$.
The solid line represents the temporal derivative of the actual energy 
as a function of time. This curve is below, but very close to,
the envelope of the curves corresponding to the individual relaxation 
processes. Note how
the combination of several simple exponential relaxation processes gives
rise to a power law relaxation.

Surprisingly, $\alpha$ is fairly independent of temperature, size,
type of interaction considered, and  localization radius,
facts for which we do not have any interpretation. Anyway, the robustness of the
exponent could be a signature of self--organized criticality. 
Similar trends have been  found in experimental measurements of the 
excess conductance of 2D samples excited far from equilibrium \cite{OP97}. In the absence of
magnetic field, the power law exponent of these measurements ranges
between 0.27 and 0.29, diminishing with the strength of the magnetic field.

Our results point to the difficulty in extracting information about the
effects of interactions from the power law exponent. Nevertheless, the type 
of interaction significantly affects the longest relaxation times.

\section{Longest relaxation time}

We also study the longest relaxation time, $\tau_2$, as a function of
temperature and the size of the sample for systems with Coulomb interactions,
with short-range interactions and for non-interacting systems. 
In Fig.\ 4 we plot $\langle \log_{10} \tau_2 \rangle$ versus the 
inverse of the temperature for the three types of interactions
mentioned, Coulomb (solid lines), short-range (dotted-dashed lines) and 
no-interactions (dashed lines).
The number of sites considered are $N= 248$, 341, 465, 744 and 899, for long
range interactions and for non-interacting systems; for short range
interactions we did  not use the two largest sizes.
$\tau_2$ increases with sample size, and thus the smallest sample
corresponds to the lowest curve, and so on. 
$\langle  \rangle$ denotes averages over site configurations.
Fluctuations in $\tau_2$ from sample to sample are very large and, as is
the case with most properties of disordered systems, one has to average 
the common logarithm of $\tau_2$, rather than $\tau_2$ itself.
The curves extend over the range of validity of the results.
The 'high' temperature limit $T_{\rm max}$ depends on the energy range 
$\Delta E$ spanned by the configurations stored. 
We choose $T_{\rm max}=0.1\Delta E$.
The low temperature limit arises from the discrete nature of the spectrum
of configurations and we take it as being equal to the 
mean energy spacing of the ten lowest energy configurations $\Delta\epsilon$.

\begin{figure}
\epsfxsize=\hsize
\begin{center}
\leavevmode
\epsfbox{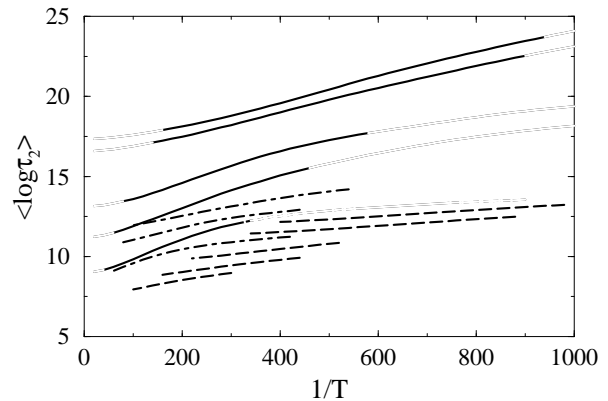}
\end{center}
\caption{Average of $\log_{10}\tau_2$ versus $1/T$ for
 systems with Coulomb interaction (solid lines), a short range
interaction (dotted-dashed lines)
and systems without interactions (dashed lines). 
The size of the systems stems from 248 sites (lowest line)
to 900 sites (highest lines). For short range interactions the largest
size considered is 465 sites.}
\label{t_log}
\end{figure}

From Fig. \ref{t_log} we can conclude that the longest relaxation time
depends strongly on the type of interaction.
$\tau_2$ is one order of magnitude larger for interacting than for 
non-interacting systems. As we will see, this effect is much larger when
extrapolated to macroscopic sizes.

\begin{figure}
\epsfxsize=\hsize
\begin{center}
\leavevmode
\epsfbox{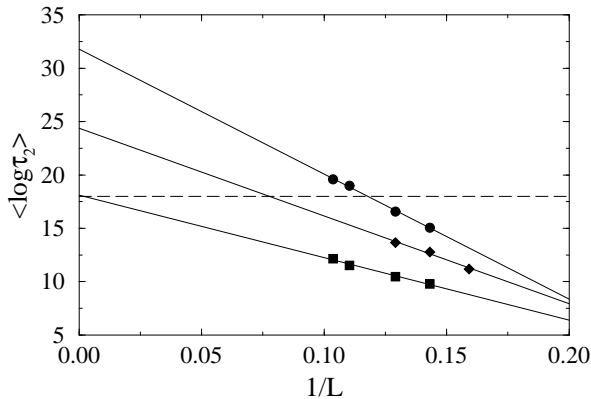}
\end{center}
\caption{Average of $\log_{10}\tau_2$ versus  $L^{-1}$ at $T^{-1}=400$ for
non-interacting systems (squares) and systems with Coulomb (dots) and
short range (diamonds) interactions.}
\label{t_L}
\end{figure}

In order to extrapolate the  previous results to macroscopic sizes we
plotted $\langle \log_{10} \tau_2 \rangle$ as a function of $L^{-\beta}$
at a fixed temperature for different values of the exponent $\beta$.
$L=N^{1/3}$ is the length 
of the side of the system, and $N$ is the number of sites. 
We found that the results for the three types of interactions fit
straight lines fairly well when $\beta=1$.
In Fig.\ \ref{t_L} we show $\log_{10}\tau_2$ versus  $L^{-1}$ for
systems with Coulomb interactions (dots), short-range interactions 
(diamonds) and without interactions (squares).
The horizontal dashed line represents a macroscopic time, say, one day 
($\approx 10^{18}\tau_0$).
The temperature chosen in this plot is $T = 0.0025$, which is valid
for the four  sizes employed in both types of interactions.
The size of the symbols used roughly corresponds to the standard 
deviation of $\log_{10} \tau_2$.
The crossing point of each straight line with the vertical axis is the 
extrapolation of $\tau_2$ to macroscopic sizes. The results are
$\tau_2^{(\infty)}\approx 10^{31\pm 1} \tau_0 = 10^{18\pm 1}$ s
(Coulomb interactions) $\tau_2^{(\infty)}\approx 10^{11\pm 1}$ s (short-range
interactions) and $\tau_2^{(\infty)}\approx 10^{5\pm 1}$ s (no interactions). 
It is clear from this figure that the longest relaxation time drastically 
increases with the strength of interactions, although these results have to 
be taken with care as they are extracted from a very long extrapolation.

The results presented in Figs.\ \ref{t_log} and \ref{t_L} correspond to
a localization radius $a=0.3l_0$. For larger values of $a$, the
relaxation times will decrease, as can be deduced from Eq.\ (\ref{omega}).
We have found empirically that a change in $a$ causes a change in $\tau_2$
of approximately $\Delta\log_{10}\tau_2\approx 3\Delta (a^{-1})$. 
The values of $\tau_2^{(\infty)}$ are so large for interacting systems that
we would  expect non--ergodic behaviour for these systems even for much larger
localization radii than the one considered here.

\section{Variable number relaxation}

At zero temperature, the relaxation process is downward in energy
and we can assume that the fastest process always dominates, 
corresponding to a well defined sequence of configurations with decreasing
energies. For each transition at $T=0$, the shorter the hopping length, 
the faster the corresponding transition rate. From each configuration, 
the system chooses the nearest one (in terms of $\sum r$) from those with less energy.
With this in mind, we have computed for all low-energy
configurations the closest one of smaller energy, and have stored the
number of electrons $n$ participating in the transition.

\begin{figure}
\epsfxsize=\hsize
\begin{center}
\leavevmode
\epsfbox{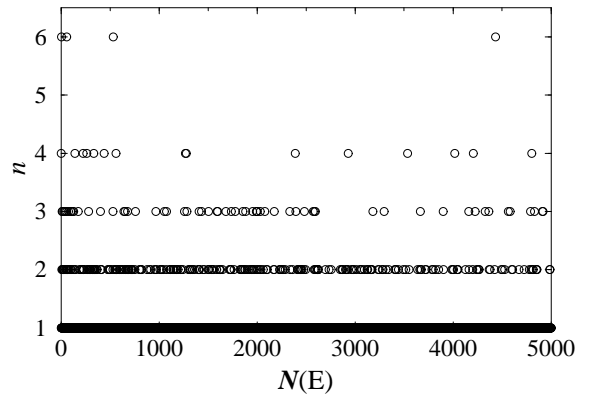}
\end{center}
\caption{Number of electrons participating in the fastest transition
as a function of the order of the initial configuration.}
\label{t_N}
\end{figure}

In Fig.\ \ref{t_N} we show the number of electrons, $n$, of the fastest
transition from an initial configuration as a function of the number
of this configuration for a Coulomb interacting sample with 900 sites.
At very low energies, the relative importance of many-electron transitions 
increases. The proportion of transitions with a fixed number of electrons 
greater than one ($n>1$) increases with decreasing energy. 
Obviously, in the non-interacting case all processes are one-electron
transitions.

\section{Conclusions}

Our numerical results of relaxation in localized electronic systems show
a power law behavior with an exponent close to 0.15 and independent of
all the parameters and type of interactions considered. At very long
times, we obtain exponential relaxation with a characteristic time that
strongly varies with size, localization radius and type of interaction.
The extrapolation of this characteristic time to macroscopic sizes
predicts values much larger than the typical experimental times,
especially for the interacting cases. The strength of interactions in
experiments performed on these systems can be deduced from their longer
relaxation times.

\section{Acknowledgments}

We would like to acknowledge Prof.\ M. Pollak for useful conversations
and a critical reading of the manuscript.
We also acknowledge 
the Direcci\'on General de Investiga\-ci\'on Cien\-t\'{\i}\-fica y 
T\'ec\-nica for financial support, project number PB
96/1118, and for  APG's grant.

%\vfill\eject


\begin{thebibliography}{99}

\bibitem{CK92} M. Ben Chorin, D. Kowal and Z. Ovadyahu, {\sl Proceedings of
HTSC
and Localization Phenomena}, eds. A. Aronov, A. Larkin and V.
Lutovinov,
World Scientific (1992) %%7

\bibitem{OP97} Z. Ovadyahu and M. Pollak, {\sl Phys.\ Rev.\ Lett.},
{\bf 79}, 459 (1997).

\bibitem{VO98} A. Vaknin, Z. Ovadyahu and M. Pollak, {\sl
Phys.\ Stat.\ Sol.} {\bf 205} 395 (1998).

\bibitem{PO85} M. Pollak  and M. Ortu\~no, {\sl Electron-electron
interactions in disordered systems}, eds. A.L. Efros and M. Pollak,
North-Holland, 287 (1985).

\bibitem{MP91} M. Mochena and M. Pollak, {\sl Phys. Rev. Lett.}, {\bf 67}
 109 (1991). %%2

\bibitem{ST93} M. Schreiber and K. Tenelsen, {\sl Europhys. Lett.}, {\bf 21}
 697 (1993).%%3

\bibitem{DM98} A. D\'\i az--S\'anchez, A.~M\" obius, M. Ortu\~no, A. P\'erez--Garrido 
and M. Schreiber, {\sl Phys.\ Stat.\ Sol.} {\bf 205}, 17, (1998).



\bibitem{SE84}   B.I. Shklovskii and A.L. Efros, {\it Electronic
Properties of Doped Semiconductors} (Springer, Heidelberg, 1984).


\bibitem{PO97} A. P\'erez--Garrido, M. Ortu\~no, E.\ Cuevas, J. Ruiz and
M. Pollak, {\sl Phys.\ Rev.\ B} {\bf 55} R8630 (1997).

\bibitem{ADMO98} A.~D\'\i az-S\'anchez, A.~M\" obius, M.~Ortu\~no,
   A.~Neklioudov, and M.~Schreiber, to be published. 

\bibitem{MOPO96} A.~M\" obius and M.~Pollak, Phys.\ Rev.\ B {\bf 53},
   16\, 197 (1996).

\bibitem{MO97} A.~M\" obius, A.~Neklioudov, A.~D\'\i az-S\'anchez,
   K.H.~Hoffmann, A.~Fachat, and M.~Schreiber, Phys.\ Rev.\ Lett.\
   {\bf 79}, 4297 (1997).


\bibitem{RC94} J. Ruiz, E. Cuevas, M. Ortu\~no,  J. Talamantes, M.
Mochena and M. Pollak, {\sl J. Non-crystalline Solids}, {\bf 172--174},
 445 (1994).

\bibitem{PO98} A. P\'erez--Garrido, M. Ortu\~no and A.~D\'\i az-S\'anchez, 
{\sl Phys.\ Stat.\ Sol.} {\bf 205} 31 (1998).


\end{thebibliography}
\end{document}